\documentclass{aa}

\usepackage{graphics}

\usepackage{natbib}
\bibpunct{(}{)}{;}{a}{}{,}

\begin{document}

\title{
	Analysis of the solar cycle and core rotation \\ 
	using 15 years of Mark-I observations:1984-1999.
	}

\subtitle{I. The solar cycle}

\author{
	S.J. Jim\'enez-Reyes\inst{1,2} \and
	T. Corbard\inst{1}\fnmsep\thanks{\emph{Present address:} Intitute of Astronomy, University of Cambridge, Madingley Road, Cambridge CB3 OHA, UK.} \and
	P.L. Pall\'e\inst{2} \and
	T. Roca	 Cort\'es\inst{2,3}\and
	S. Tomczyk\inst{1}
	}

\offprints{S.J. Jim\'enez-Reyes, \email{chano@ucar.edu}}

\institute{
	High Altitude Observatory, NCAR, PO Box 3000, 
		Boulder, CO 80307 USA \and
	Instituto de Astrof\'\i sica de Canarias, E-38701, 
		La Laguna, Tenerife, Spain	\and	
	Departamento de Astrof\'\i sica, Universidad de La Laguna, 
		Tenerife, Spain
	}

\date{Received March 30, 2001; accepted October 3, 2001}

\abstract{
High quality observations of the low-degree acoustic modes ($p$-modes)
 exist for
almost two complete solar cycles using the solar spectrophotometer
Mark-I, located  at the Observatorio del Teide (Tenerife, Spain) 
and operating now as part of the Birmingham Solar Oscillations 
Network (BiSON). We have performed a Fourier analysis of 30 calibrated 
time-series of one year duration covering a total period of 15 years 
between 1984 and 1999. Applying different techniques to the resulting 
power spectra, we study the signature of the solar activity changes on 
the low-degree $p$-modes. We show that the variation of the central 
frequencies and the total velocity power (TVP) changes. 
A new method of simultaneous fit is developed and a 
special effort has been made to study the frequency-dependence
of the frequency shift. The results confirm a variation of the 
central frequencies of acoustic modes of about 0.45~$\mu$Hz, 
peak-to-peak, on average for low degree modes between 2.5 and 3.7~mHz. 
The TVP is anti-correlated with the 
common activity indices with a decrease of about 20$\%$ between the
minimum and the maximum of solar cycle 22. The results are compared 
with those obtained for intermediate degrees, using the LOWL data. 
The frequency shift is found to increase with the degree with 
a weak $\ell$-dependence similar to that of the inverse mode mass. 
This verifies earlier suggestions that near surface effects are predominant.
\keywords{Sun: activity -- Sun: oscillations -- Sun: interior -- 
		Sun: rotation -- Methods: data analysis
	}
}

\authorrunning{Jim\'enez-Reyes et al.}
\titlerunning{Analysis of the solar cycle and core rotation. I}
\maketitle

\section{Introduction}
Understanding the  observed solar variability is one of the major goals
of solar physics. Because the frequency shifts of solar $p$-modes are
known to be very sensitive to the solar activity cycle, the analysis of
helioseismic data has been used to track those physical processes
which underly the origin of the cyclic changes observed at the solar surface. 
Helioseismology based on low-degree $p$-modes is necessary to look 
for potential structure or dynamic changes in the deep interior.

The first report of frequency shifts of the low-degree $p$-modes 
was given by \citet{woodard85}. Using ACRIM data, they found that the few 
observed $\ell$=0 and $1$ modes presented a change in the central frequency 
of 0.42$\pm$0.14~$\mu$Hz in average during the declining phase of cycle 21 
(1980-1984). These results were confirmed by \citet{fossat87} by 
comparison with observations made at the south pole, 
and by \citet{palleetal89} using a long set of data from the 
Mark-I instrument at 
Observatorio del Teide covering the full cycle 21 (1977-1988). 

Subsequently, \citet{regulo94} used
Doppler observations collected from the maximum of cycle 21 to the 
falling phase of cycle 22 (1980-1993), obtained with Mark-I instrument, 
to calculate monthly  frequency shifts. They showed that, for all 
low degree acoustic modes, there is an important frequency shift of 
0.52$\pm$0.02~$\mu$Hz correlated with solar activity. 
In addition, the amplitude of these variations is different 
when $\ell$=1,3 and $\ell$=0,2 are considered separately. 
The odd modes present, on average, a change of 0.58$\pm$0.06~$\mu$Hz;
the even ones show a full shift of only 0.33$\pm$0.06~$\mu$Hz.

Two other important properties of the low-degree changes have also been 
pointed out recently. \citet{gubau92} observed  the frequency dependence 
of the frequency shifts for the low-degree $p$-modes, 
\citep[later confirmed by][]{chaplin98a} in agreement with 
the earlier results of 
\citet{woodard91} for intermediate degrees. \citet{jimenez-reyes98} 
found that the frequency shifts, when plotted against an activity index, 
show a hysteresis behavior rather than a simple linear correlation.
This result was interpreted as part of structural changes associated with 
the solar activity which are taking place in the Sun. This must be 
confirmed by more observations but the interpretation of these later 
results as being partly due to structural changes in the interior
associated with the solar activity has been found to be a complex problem. 
Recently, \citet{fmi00} have studied in detail the signature left on the 
low-degree $p$-mode frequencies by the surface solar  magnetic activity.
Whether these changes are taking place only close to the surface or not 
is not completely clear and one of the requirements 
to address this question is to get precise and reliable 
measures of the low degree mode parameters for a 
long period of time. 

The spectrophotomer Mark-I, has been collecting solar observations for almost 
two complete solar cycles. The available database for low-degree $p$-modes, 
probably the longest in duration and the most stable, is used in the 
present work to analyze the signature of the solar cycle in the mode 
parameters and to parameterize  the observed
frequency shifts as a function of various classical solar indices.

In the following section, the essential steps of the data reduction 
leading to the yearly spectra
are presented.  In general, the frequency shift between a time $t_i$ 
and a time $t_o$ taken as reference, can be written as a function of the  
frequency and the degree, i.e. $\delta\nu(t-t_o,\nu,\ell)$.
In the following the reference time is $1986$ which corresponds to a 
minimum of solar activity. In Sect.~\ref{sec:int_dnu}, the integrated
frequency shift ${\Delta\nu}^i\equiv<\delta\nu(t_i-t_o,\nu,\ell)>$
is analyzed where the brackett indicates an average for all observed low degrees and frequencies between 
2.5 and 3.7~mHz. Two techniques are proposed to measure the frequency
shift from the cross-correlation function between power
spectra of time-series created at different solar activity
level. In addition to the frequency shift, 
the second technique   allows us to study the
time variation of the total velocity power (TVP) which are presented in 
Sect.~\ref{sec:tvp}.
Then, in Sect.~\ref{sec:nu-dep}, we focus on the study of the 
frequency dependence of the frequency shifts. Again, two different techniques 
are used.  The first one consist in simply cutting the spectra in band
of 135~$~\mu$Hz  before computing  the cross-correlation functions. The 
second one is a  new procedure developed here and 
called simultaneous fitting:
all the yearly spectrum are fitted at the same time assuming that the
time dependence of the mode frequencies  
can be described as a linear function of  the radio flux 
 at 10.7 cm $F_{10}^i$, taken as solar activity index i.e.:
\begin{equation}\label{eq:def1}
\delta\nu(t_i-t_o,\nu,\ell)=\delta\nu(\nu,\ell)(F_{10}^i-F_{10}^o),
\end{equation}
where $F_{10}^o$ represents the radio flux at the 1986 solar minimum.
In addition, we define $\delta\nu(\nu)$ as the frequency shift per radio 
flux unit averaged over $\ell$. 
We note that if we had used shorter time-series and a magnetic index
instead of the radio
flux  to parameterize the time
dependence of the frequency shift, a more complicated formulation would
probably have been
needed in order to take into account  the hysteresis behaviour found
when magnetic indices  are plotted versus the frequency shift during the
cycle \citep{jimenez-reyes98}.

In order to check the $\ell$-dependence of the frequency shift, we use the fact that 
pairs of low-degree $p$-modes with the same parity have ``almost'' the same 
frequency and, that they are equally spaced in frequency. This allows us 
to provide in Sects.~\ref{sec:int_dnu}, \ref{sec:tvp}, \ref{sec:nu-dep}
not only the $\ell$-averaged quantities 
defined above ($\Delta\nu$, TVP, $\delta\nu(\nu)$) 
but also the quantities related to even and odd  modes separately i.e. 
respectively: $\Delta\nu_{0,2}$, TVP$_{0,2}$, $\delta\nu(\nu)_{0,2}$ 
and $\Delta\nu_{1,3}$, TVP$_{1,3}$, $\delta\nu(\nu)_{1,3}$.
Finally, in Sect.~\ref{sec:l-dep} the $\ell$-dependence analysis of 
the frequency shift is completed by comparing the results with those 
obtained at higher degrees ($\ell=1,99$) using the LOWL database.

\section{Observations and data analysis}\label{sec:obs}
The data used in this work come from the observations carried out at 
the Observatorio del Teide between 1984 and 1999. 
The observations consist of daily measurements of the solar radial velocity 
obtained with the Mark-I resonant scattering spectrophotometer. 
This instrument has been sited at the Observatorio del Teide
since 1975. After some hardware updates in 1984, the experiment has been 
running without interruption other than bad weather and instrumental
failures. The data reduction process is explained in more
detail elsewhere \citep{vdraay85,palle86,palle93}.
Briefly, the data are corrected from the annual scan 
(Earth's orbit around the Sun) 
of the non-linear solar line shape, and calibrated by 
fitting the known daily velocity of the observatory;
only two parameters are fitted, taken to be the same over the all 15 
years analyzed here. Then, the daily residuals are joined in 
consecutive 360 days leading
to a total of 30 time-series with 6 months in common between
consecutive series. Although all series are not independent, they show very 
similar duty cycles (around $25\%$).
Finally, the corresponding power spectra were calculated for every 
time-series using a traditional Fourier analysis. 
All the spectra show the peaks of the low-degree $p$-modes with $\ell\leq3$. 
The typical sideband structure appears in the spectra at $k_D$=11.57~$\mu$Hz
as a direct consequence of the observing window achieved 
from just one station. 

\section{Frequency integrated frequency shifts}\label{sec:int_dnu}

 Once the power spectrum of each time-series has been calculated,
the integrated frequency shifts $\Delta\nu^i$ are  determined. 
This is done by computing the cross-correlation 
$\rho^i(\nu_{j})$ of each  power spectrum  $i$ with the power spectrum
of the time-series covering 1986 taken as reference. 
As we show in Sect.~5, the  dependence of the frequency shift for low 
and intermediate-degree $p$-modes appears to be essentially 
null below 2~mHz, whereas at high frequency (above 3.7~mHz) the 
frequency shift is expected to drop quickly. The analysis of the
integrated changes is limited to the modes between 2.5 and 3.7~mHz. 
considering a larger interval  could 
bias the results by including  the  strong high frequency variations 
that are less accurately determined and 
have possibly different physical explanation 
\citep{goldreich91}.

In order to calculate the position of the cross-correlation main peak, 
two different methods have been used. The first was used in previous 
analyses \citep{regulo94,jimenez-reyes98} 
and takes the maximum of a  second-order polynomial fitted 
to the logarithm of the cross-correlation function in an 
interval $\pm\sigma$ around the main peak (where $\sigma$ is 
the second-order moment). This cross-correlation is calculated 
starting at an appropriate lag for which the function is symmetric, 
and which is obtained by calculating the third-order moment. 
This procedure directly provides a value of the mean frequency shift between
the $\ell\leq3$ $p$-modes and  the corresponding values at solar activity 
minimum, the chosen reference.

The second method, introduced here, is based on the shape of 
the cross-correlation function.
Assuming that each oscillation mode can be modeled by a damped
harmonic oscillator, each peak in the spectra has a Lorentzian profile and
the correlation function has also a Lorentzian profile. 
 In order to  improve the determination 
of the frequency shift, the model also take into account the presence
of sidebands with amplitude $\beta_{|1|}$ located at $\pm$11.57~$\mu$Hz.
 Thus, our model  of the cross-correlation function between the  spectrum $i$ 
and the reference spectrum can be written as:
\begin{equation}
\label{mcross}
M^i_\rho(\nu_j)= \!\!\!\!\sum_{k=-1}^{1} \! \beta_{|k|} \frac{ A^i
(\Gamma^i/2)^2}{(\nu_j-\Delta\nu^i+k_D
\cdot k)^2+ (\Gamma^i/2)^2} +B^i,
\end{equation}
where the parameters to be fitted are:
\begin{itemize}
\item $A^i$, the amplitude of the central peak;
\item $\Gamma^i$, the linewidth of the Lorentzian profile;
\item $\Delta\nu^i$, the average frequency shift in the chosen frequency
                 interval; 
\item $B^i$, the constant background level;
\item $\beta_{|k|}$,  the ratio of the sidebands to the central peak 
                  ($\beta$=1 for $k$=0 );  
\item $k_D=11.57\mu$Hz is the constant separation 
of the sidebands and it is the only fixed parameter 
in the fit.
\end{itemize}
 The best estimation, in the least square sense,
of the parameters related to each time-series $i$, 
is obtained by minimizing the following quantities: 
\begin{equation}
\label{fitccf}
\chi^{2}_i = \sum_{j=1}^{N}{\mid \rho^i(\nu_{j}) - M^i_\rho(\nu_{j}) \mid^{2}},
\end{equation} 
where $N$ is the total number of frequency bins in an interval 
of $\pm$20~$\mu$Hz around the main peak 
of the cross-correlation function. We applied a Levenberg-Marquard 
method (Press et al. 1992) but any other minimization routine may be used.
 
\begin{figure}
\resizebox{\hsize}{!}{\includegraphics{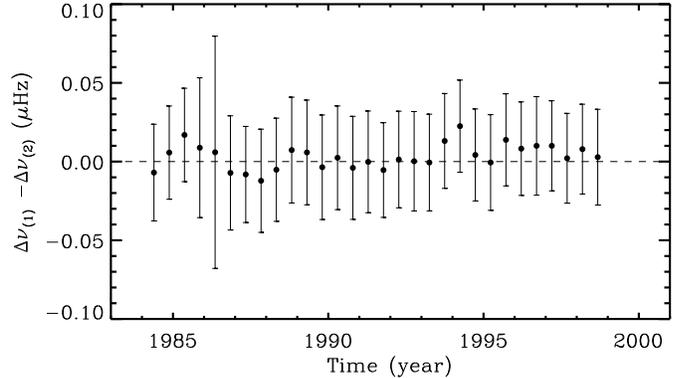}}
\caption{Differences between the integrated frequency shifts as inferred
from the cross-correlation functions by the two methods explained in the text 
i.e. (1) fit by a polynomial close to the maximum and (2) fit by a Lorentzian profile including side bands. The differences remain small and not significant.}
\label{fig:diff_fit_ccf}
\end{figure}

We note that the first technique is more objective in the sense that it 
does not require  a physical hypothesis or modeling of the $p$-mode 
excitation and damping mechanisms.
It uses only information contained in the spectrum while the second method 
assumes symmetric Lorentzian profiles. There is some evidence that 
the shape of  the peaks in the power spectra are slightly asymmetric 
\citep{sabri00}.  
In order to check the efficiency of both methods and the
influence of the duty cycle in the final result, we analyzed the 
frequency shifts in periods of 36 days for which the duty cycle varies.
The differences between the two  methods remain in general within the error
bars. The only significant differences are found
for time-series with very small duty cycle (around $10\%$) 
but, for the yearly time-series analysed here, the duty cycle 
 is moderately high and quite stable (around $25\%$)
from year to year. Regarding to the observed asymmetry of the
$p$-modes, they are not thought to be important for this analysis. 
The second method takes into account the known distance from the 
sidebands to the main peak ($\pm$11.57~$\mu$Hz) which  improve 
the results. Moreover this method allows 
to study  not only the frequency shift but also the TVP by providing 
amplitudes and
widths of the cross-correlation functions (see Sect.~\ref{sec:tvp}).
Therefore hereafter only the second method will 
be considered but, for completeness, 
Fig.~\ref{fig:diff_fit_ccf} shows the marginal 
differences between the frequency shifts obtained
by applying both  methods to the yearly time-series.

\begin{figure}
\resizebox{\hsize}{!}{\includegraphics{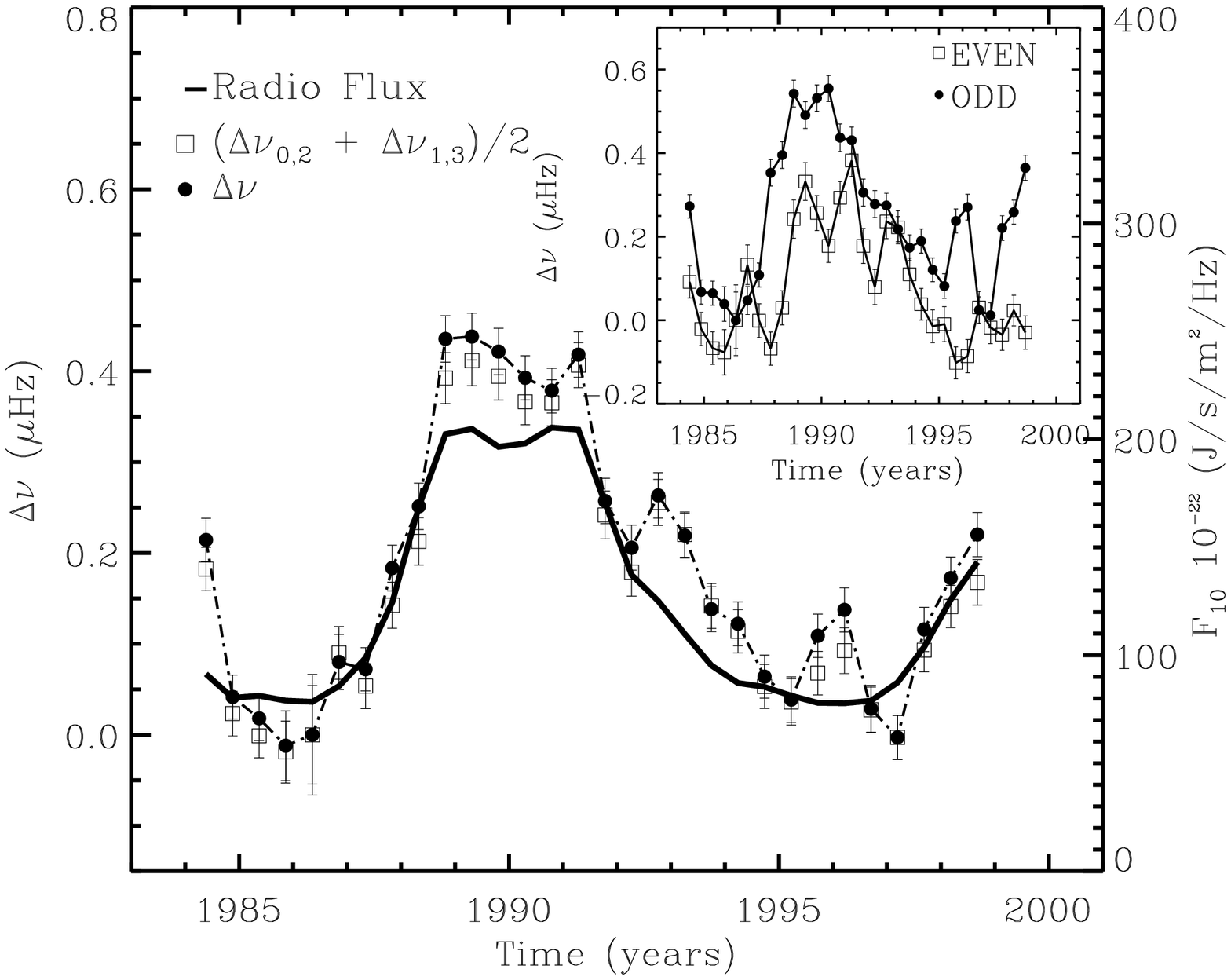}}
\resizebox{8cm}{!}{\includegraphics{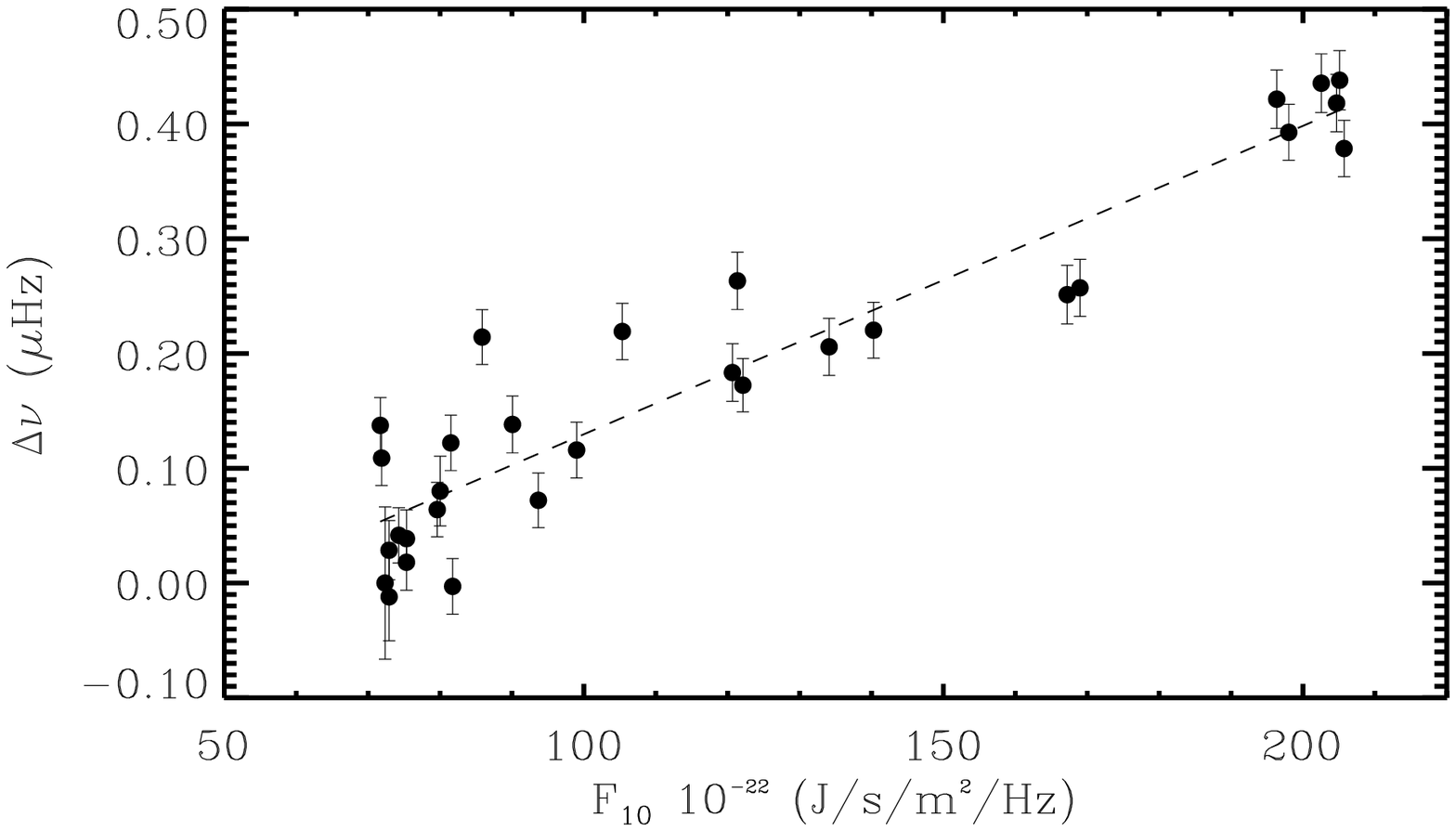}}
\caption{Time variation of the integrated frequency 
shift for low-degree $p$-modes is plotted in the upper panel (black dots), 
where the radio flux at 10.7~cm is also shown (full line).
In the sub plot, the results for the even and odd degrees separately 
are presented. The averages of these, are plotted in the main panel (squares).
The lower figure shows the good linear correlation between the frequency
shift and the radio flux.}
\label{fig:dnui}
\end{figure}

\begin{table*}
\begin{center}
\caption{Correlation coefficients between different solar indices 
and the yearly frequency shift. $r_P$ is the Pearson linear correlation 
coefficient, $r_S$ the Spearman rank correlation coefficient and $P_s$ 
is the probability of having no correlation.}
\begin{tabular}{lccccccccccc} 
\hline
Index & & $\Delta\nu$ & & & & $\Delta\nu_{0,2}$ & & & & $\Delta\nu_{1,3}$ \\ 
\cline{2-4}\cline{6-8}\cline{10-12}
& $r_P$ & $r_S$ & $P_s$ & &  $r_P$ & $r_S$ & $P_s$ & & $r_P$ & $r_S$ & $P_s$  \\
\hline
R$_{I}$ & 0.94 & 0.88 & 9 10$^{-11}$ 
& &  0.76 & 0.70 &  1 10$^{-5}$ 
& &  0.90 & 0.85 & 4 10$^{-9}$ \\

F$_{10}$& 0.94 & 0.87 & 4 10$^{-10}$ 
& &  0.77 & 0.72 & 7 10$^{-6}$ 
& &  0.90 & 0.83 & 1 10$^{-8}$ \\

KPMI & 0.90 & 0.86 & 8 10$^{-10}$ 
& & 0.79 & 0.71 & 1 10$^{-5}$ 
& & 0.85 & 0.84 & 8 10$^{-9}$ \\

MPSI & 0.94 & 0.88 & 5 10$^{-10}$ 
& & 0.83 & 0.76 & 2 10$^{-6}$ 
& & 0.88 & 0.85 & 1 10$^{-8}$  \\

TSI& 0.89 & 0.89 & 3 10$^{-8}$ 
& & 0.78 & 0.78 & 1 10$^{-5}$ 
& & 0.83 & 0.83 & 2 10$^{-7}$ \\

He& 0.94 & 0.88 & 9 10$^{-11}$ 
& & 0.79 & 0.74 & 2 10$^{-6}$ 
& & 0.89 & 0.83 & 2 10$^{-8}$ \\
\hline
\end{tabular}
\label{corr_dnu}
\end{center}
\end{table*}

As mentioned in the introduction, the asymptotic theory predicts that, for low-degree $p$-modes, pairs of modes 
with alternately odd and even degrees are equally spaced in frequency 
with a separation of about  $67$~$\mu$Hz i.e. half of the so-called big 
separation \citep[e.g.][]{Deubner84}.
The contributions of even ($\ell$=0,2) and odd degrees 
($\ell$=1,3) to the integrated frequency shift can therefore
be separated simply by applying a mask to the spectra before
using the procedure explained above. 
It should also be noted that, for full disk observations, the contribution 
of the two modes of a pair is not the same, due to the geometry of the 
modes at the surface.
In the case of the odd modes, the frequency shifts come essentially 
from $\ell$=1 due to the high ratio in sensitivity 
between $\ell$=3 and 1 ($\sim$ 0.1) whereas, 
in the case of the even modes the sensitivity ratio is close to one 
and therefore the contributions of $\ell$=0 and 2 are nearly the same. 

Figure~\ref{fig:dnui} illustrates the frequency averaged frequency shift 
between $2.5$ and $3.7$~mHz. 
The results corresponding to $\ell$=0,2 and $\ell$=1,3 
have been plotted in the sub plot at the top-right corner and their average 
is shown in the main figure. The solid line represents the radio flux 
averaged over the same periods as the time-series used for 
this work. 
 
Although there are few departures from the general trend which do not 
agree with the smooth behavior of the solar index, the integrated 
signal concurs very well with the radio flux, which
represents here the behavior of the solar cycle. 
The average of both contributions ($\ell$=0,2 and  $\ell$=1,3) follows well 
the integrated signal as expected.
The amplitude, measured as the straight difference from peak-to-peak, 
for all observed modes is $0.45\pm0.05$~$\mu$Hz,
while in the case of the even and odd degree modes are 0.48$\pm$0.05
and $0.55\pm0.07$~$\mu$Hz respectively. 
Although the difference in amplitude between even and odd degree modes 
seems to remain, the frequency shift for the even modes is larger than that 
obtained by \citet{regulo94}.

\begin{table}
\begin{center}
\caption{Intercept and slope of the frequency shift expressed as a linear
function of different solar indices.}
\begin{tabular}{llcc} 
\hline
Solar& & Intercept $a$ & Slope $b$\\
Index& & (nHz)     & (nHz per activity unit$^*$) \\
\hline
R$_I$ & 
$\Delta\nu$      & 18.25$\pm$14.77 & 2.56$\pm$0.18 \\
& $\Delta\nu_{0,2}$&-48.08$\pm$25.94 & 1.98$\pm$0.32 \\
& $\Delta\nu_{1,3}$& 60.62$\pm$21.51 & 2.91$\pm$0.26 \\
\hline
F$_{10}$ &
$\Delta\nu$      &-139.07$\pm$24.17 & 2.69$\pm$0.19\\
& $\Delta\nu_{0,2}$&-173.73$\pm$41.78 & 2.11$\pm$0.32\\
& $\Delta\nu_{1,3}$&-115.88$\pm$36.40 & 3.03$\pm$0.28\\
\hline
KPMI & 
 $\Delta\nu$      & -98.43$\pm$27.42 & 22.92$\pm$2.04\\
& $\Delta\nu_{0,2}$&-154.75$\pm$37.20 & 19.08$\pm$2.78\\
& $\Delta\nu_{1,3}$& -63.15$\pm$40.42 & 25.34$\pm$3.02\\
\hline
MPSI & 
$\Delta\nu$      & 33.79$\pm$14.04 & 149.06$\pm$10.46\\
& $\Delta\nu_{0,2}$&-39.06$\pm$21.10 & 124.22$\pm$15.59\\
& $\Delta\nu_{1,3}$& 79.81$\pm$22.19 & 164.63$\pm$16.56\\
\hline
TSI & 
$\Delta\nu$        &-528.06$\pm$51.63 & 386.71$\pm$37.80\\
& $\Delta\nu_{0,2}$&-437.13$\pm$65.43 & 320.07$\pm$47.90\\
& $\Delta\nu_{1,3}$&-585.31$\pm$71.58 & 428.67$\pm$52.40\\
\hline
He &
$\Delta\nu$      &-438.86$\pm$42.45 & 10.92$\pm$0.73\\
& $\Delta\nu_{0,2}$&-417.59$\pm$73.56 &  8.87$\pm$1.26\\
& $\Delta\nu_{1,3}$&-447.34$\pm$67.87 & 12.21$\pm$1.17\\
\hline
\end{tabular}
\label{slope_dnu}
\end{center}
{\small
$^*$Units: 
nHz; nHz/(10$^{-22}$ J/s/m$^2$/Hz), nHz G$^{-1}$, nHz G$^{-1}$,
nHz W$^{-1}$m$^2$, nHz m\AA$^{-1}$ respectively.}
\end{table}

Since the time-series created are one year long, the time variation 
of the integrated frequency shift analyzed here informs  only on long-term 
changes. A  recent analysis covering different
time scales from one to seven months using 9 years of BiSON data
can be found in \citet{Chaplin2001}. 
Average values 
of the following solar activity indices have been computed over the same
one year periods  than the frequency shifts in order to obtain the 
corresponding correlation coefficients:


\begin{itemize}
\item the International Sunspot Number, R${_I}$;
\item the integrated radio flux at 10.7 cm, F$_{10}$; \\
(both obtained from the Solar Geophysical Data)\footnote{Available at http://web.ngdc.noaa.gov/stp/stp.html};
\item  the Kitt Peak magnetic index (KPMI) extracted from the Kitt Peak full
disk  magnetograms \citep{harvey84};
\item the Mount Wilson Magnetic Plage Strength Index 
\citep{ulrich91}, MPSI;
\item the Total Solar Irradiance, TSI \citep{frohlich98}; and
\item the equivalent width of HeI 10830\AA \ averaged over the 
whole solar disk using data from Kitt Peak observatory.
\end{itemize}
 
The Pearson correlation coefficient $r_P$, which is 
a measure of the strength of the linear relationship between two 
indices, and the Spearman rank correlation coefficient, $r_S$ 
which provides a measure of the correlation between the ranks of two indices 
during the chosen period, are shown in Table~\ref{corr_dnu} for
the frequency shift.  The correlation
between the different solar activity indices themselves 
have been investigated in details by \citet{BW94}.

In addition,  the probability $P_s$ of  having null correlation between 
the ranks of any of the solar indices and the frequency shift is indicated. 
We note that the database of MPSI and TSI indices, 
covers only the first 28 time-series. The correlation analysis for 
these two indices was therefore made with two fewer points than for 
the others.

The integrated signal shows very high correlation with the various 
solar indices, whereas the frequency shift corresponding
to $\ell$=1,3 and, more significantly, $\ell$=0,2 are slightly lower. 
The general trend of the frequency shift corresponding to
the even and odd degree modes separately is quite clear but, 
in addition to the difference in amplitude already mentioned,
the frequency shifts measured for the even modes seem 
to be sensitive to the solar cycle later than the odd ones 
and the resulting difference in phase between the two curves 
is probably at the origin of the lower correlation 
coefficients found for $\ell$=0,2.
As demonstrated by \citet{fmi00} different modes would respond differently
at different phases of the cycle depending on the positions of the activity
(i.e. sunspots) on the disk.
Aside of the long term differences, there are also fluctuations
at shorter time scales which are different for the two data sets.

Because of the excellent linear correlation coefficients found,
the frequency shifts were fitted as a linear function of 
the different solar indices $I$ by: 
\begin{equation}\label{eq:fit}
\Delta\nu^i=a + b \cdot I^i.
\end{equation} 
In Table~\ref{slope_dnu} we report
the intercepts and the slopes obtained for all the solar indices
considered, and an example is given for the radio flux in the lower part of
Figure~\ref{fig:dnui}. 
 The slopes can be compared with the results shown
in Table~1 of \citet{regulo94} also obtained for low-degree $p$-modes.
The three solar indices in common to both works
present similar slopes; the differences are less than 3 times
our error bars. 
Notice that, more or less, all activity indices used here produce
similar values of the correlation coefficient with the frequency 
shifts leading to the conclusion than none is much better, for this purpose, 
than others.

The numbers also agree with those obtained recently by
\citet{jain00} for intermediate-degrees ($\ell$=20-100) using GONG data. 
However, intermediate degree modes are confined closer to the surface
and one may expect them to be more sensitive to the activity changes and 
therefore the slope to be larger for those modes than for low-degree modes.
This is indeed what we found analyzing LOWL data \citep{jimenez-reyes01a}
but, as we shall see, this depends also strongly on the range of 
frequencies considered to calculate the average values.

The interpretation of the different behavior found for $\ell=$1,3 
and $\ell$=0,2 is not straightforward. In order to 
understand better the underlying physics, one may  instead look at 
the velocity power variations and at the frequency and $\ell$-dependences. 
This is considered in the following section.

\begin{figure}
\resizebox{\hsize}{!}{\includegraphics{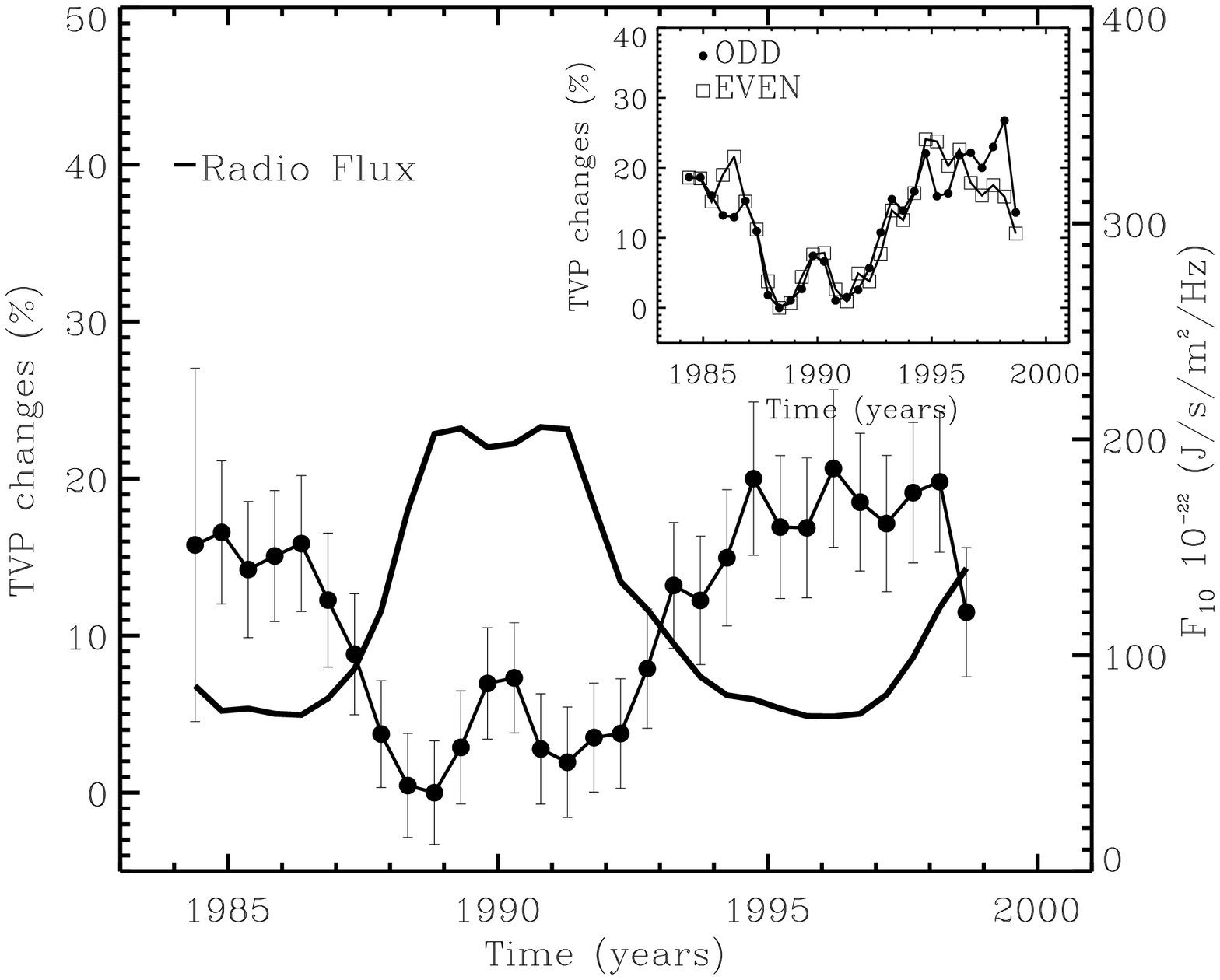}}
\resizebox{8cm}{!}{\includegraphics{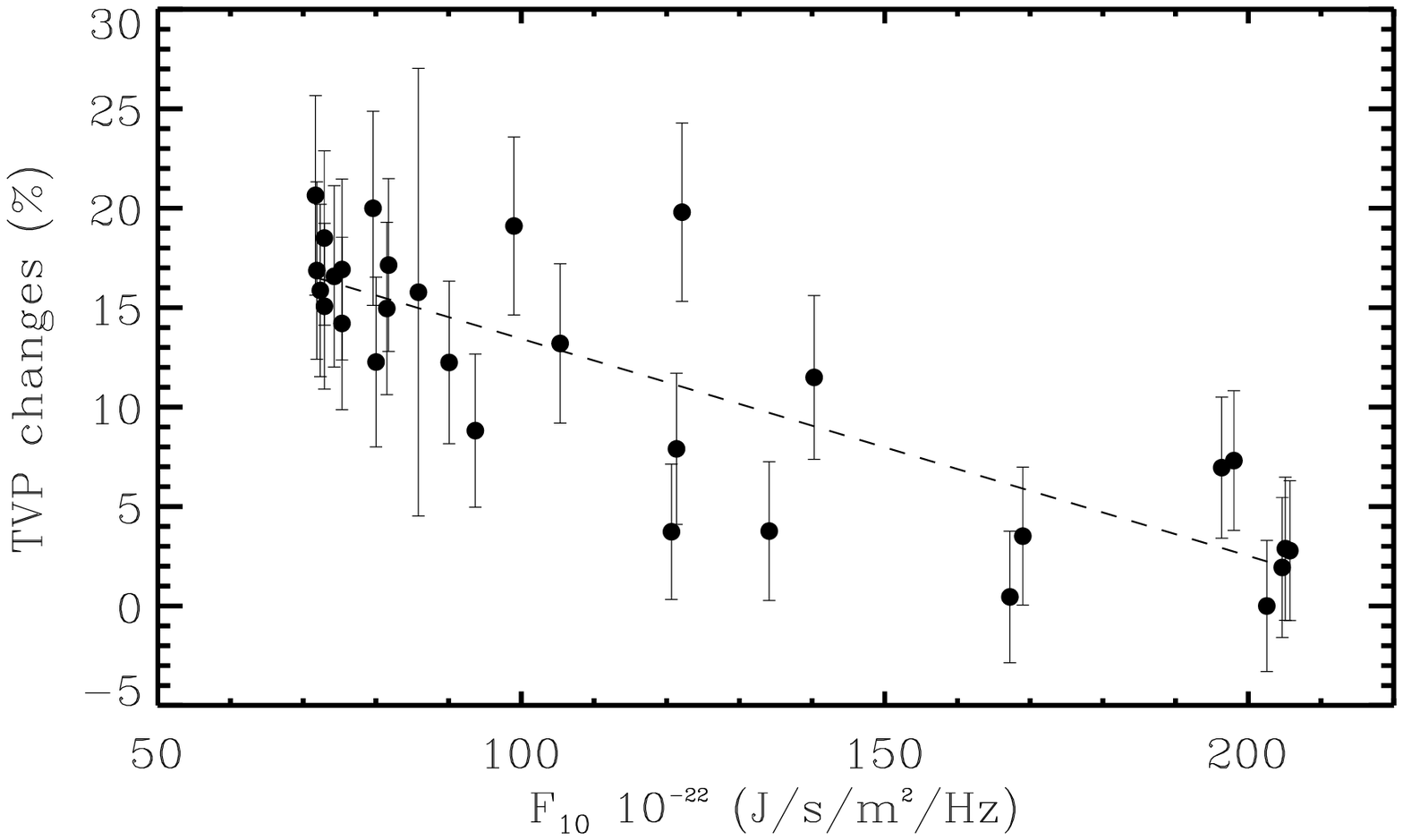}}
\caption{ Time variation of the TVP changes in percent compare to its minimum value (i.e. $100\cdot(TVP^i-TVP_{min})/TVP_{min}$) for low-degree
$p$-modes integrated between $2.5$ and $3.7$~mHz. In the sub-panel the 
TVP corresponding to $\ell$=1,3 and $\ell$=0,2 are also shown.
The radio flux at 10.7 cm calculated for the same periods is plotted as a 
full line in the upper plot and its linear correlation with the 
frequency shift is shown in the lower plot.}
\label{dos}
\end{figure}
\begin{figure}
\resizebox{\hsize}{!}{\includegraphics{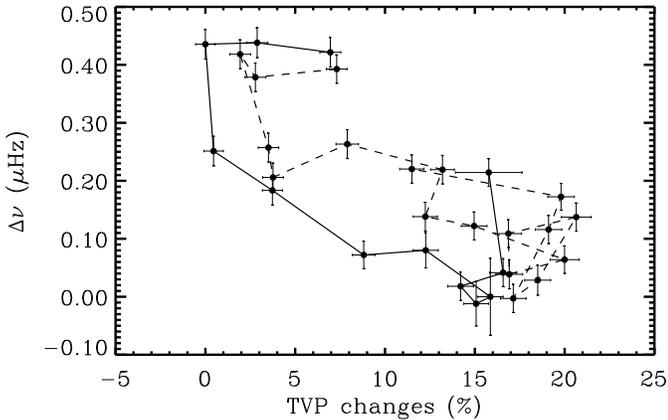}}
\caption{TVP increase in percent against the frequency
shift. The solid line goes from 1984 to 1989 and the dashed line goes from
1990 to 1999. The corresponding correlation coefficients are shown in Table~\ref{corr_ener}.}
\label{fig:dnu_tvp}
\end{figure}

\section{Total velocity power variations}\label{sec:tvp}

The first observations of the variation  of the TVP
for all measured $p$-modes was reported by  \citet{palle90a,palle90b} who 
found an increase of 30 to $40\%$, anti-correlated with the solar 
activity cycle. 
Afterwards, \citet{gubau92} found similar changes using different analysis
techniques and more data of the same type; 
they interpreted these results as a decrease in the efficiency 
of the excitation of such modes at solar activity maximum, 
since absorption of mode power by local magnetic structures 
\citep[see e.g.][]{bogdan93} is a small influence and cannot 
explain such a high ratio.

We have also calculated the TVP
in the spectrum which is proportional to the area under the main peak 
of the cross-correlation function. Once the main peak is fitted to a 
Lorentzian profile, the TVP 
is calculated as the amplitude times the width (i.e $TVP^i=A^i\cdot \Gamma^i$).
We show in  Fig.~\ref{dos}, the percentile 
changes of the TVP compared to its minimum value.
 Again, the radio flux is shown here as an index
of the solar activity cycle. From the figure, we are able to see that the
variation of the TVP between minimum and maximum
of the solar cycle is around 20$\%$.
Moreover, the lower figure shows clearly the anti-correlation
with the solar cycle, the TVP decreases when activity increases. 
 The separated contributions of the odd and even
pairs of modes were also studied (see the inset box in Fig.~\ref{dos}) 
as we did in the analysis 
of the frequency shifts. The results are shown in the sub panel 
on Fig.~\ref{dos}. Notice that the variation of the TVP between 
extreme phases of the cycle for even and odd degrees is approximately 20$\%$,
in good agreement with the amplitude of the integrated measurement.
Moreover, both contributions are in phase,
which contrasts with the results found for the frequency shifts,
where even modes seems to respond later than the odd ones. 
Odd and even TVP appear also better correlated around
the maximum than during the low activity phases: both
present exactly the same bump in the middle of the
maximum of activity close to 1990, anti-correlated with
decreasing activity during the same period, 
but their short term variations are different around 1986
and 1996 during the activity minima

\begin{table*}
\begin{center}
\caption{Correlation coefficients between the TVP and the solar
indices where the TVP is defined as the product between the width and amplitude of the cross-correlation function between a spectrum and the 1986 reference spectrum. TVP refers to the average over the whole observed $\ell$ range ($\ell=0,1,2,3$) whereas TVP$_{0,2}$ and TVP$_{1,3}$ refer to the average for even and odd degrees respectively, all quantities being frequency integrated values between 2.5 and 3.7~mHz. The last line gives the correlation coefficients between the TVP 
and frequency shift. $r_P$ is the Pearson linear correlation 
coefficient, $r_S$ the Spearman rank correlation coefficient and $P_s$ 
is the probability of having no correlation.}
\begin{tabular}{lccccccccccc} 
\hline
Index & & TVP & & & & TVP$_{0,2}$ & & & & TVP$_{1,3}$ \\ 
\cline{2-4}\cline{6-8}\cline{10-12}
& $r_P$ & $r_S$ & $P_s$ & & $r_P$ & $r_S$ & $P_s$ &  & $r_P$ & $r_S$ & $P_s$ \\
\hline
$R_{I}$ & -0.84 & -0.82 & 4 10$^{-8}$   
&&  -0.87 & -0.88 & 1 10$^{-10}$  
&&  -0.79 & -0.71 & 1 10$^{-5}$  \\

$F_{10}$ & -0.83 & -0.79 & 2 10$^{-8}$ 
&& -0.86 & -0.87 & 5 10$^{-10}$  
&& -0.78 & -0.69 & 3 10$^{-5}$   \\

KPMI&  -0.78 & -0.75 & 2 10$^{-6}$
&& -0.81 & -0.81 & 7 10$^{-8}$ 
&& -0.74 & -0.67 & 6 10$^{-5}$  \\

MPSI& -0.82 & -0.80 & 3 10$^{-7}$
&& -0.83 & -0.85 & 9 10$^{-9}$ 
&& -0.80 & -0.69 & 5 10$^{-5}$ \\

TSI& -0.76 & -0.67 & 1 10$^{-4}$
&& -0.80 & -0.76 & 2 10$^{-6}$ 
&& -0.70 & -0.58 & 1 10$^{-3}$   \\

He&  -0.84 & -0.82 & 2 10$^{-8}$ 
&& -0.87 & -0.87 & 4 10$^{-10}$   
&&  -0.77 & -0.72 & 8 10$^{-6}$   \\

$\Delta\nu$ & -0.75 & -0.73 & 4 10$^{-6}$
& && &
& && & \\
\hline
\end{tabular}
\label{corr_ener}
\end{center}
\end{table*}

Linear and rank correlation coefficients were calculated with the same 
activity indicators than before and they are shown in Table~\ref{corr_ener}. 
The values found are large, showing a good correlation, but they 
are systematically lower than those for the frequency shift 
(range [0.67-0.84] against [0.86-0.94]).
Moreover, the correlations are bigger for $\ell$=0,2 than for $\ell$=1,3
contrary to the results found for the frequency shifts. 
The linear correlation coefficient between frequency shift and 
TVP change ($r_P$=-0.75, see also Table.~\ref{corr_ener}) 
is well below the $0.9$ reached in average between frequency shift and 
the solar activity indices suggesting that they 
are not linearly correlated and
Fig.~\ref{fig:dnu_tvp} shows indeed that they tend to follow an hysteresis
shape, rather than a strict line, when plotted against each other.   

This is important because if the decrease in TVP
is due to the presence of local or surface activity, as we believe 
is the cause for the frequency shift, then they should be well 
correlated with almost no hysteresis.
Therefore, the fact that the TVP is less well
correlated with the surface activity indices and shows an hysteresis 
behavior when plotted against the frequency shift, may indicate that indeed
its variation is due to 
a decrease in the excitation efficiency or an increase of the damping rate
at maximum which  reflects a change in the convection 
zone structure that does not have to be correlated or strictly in phase 
with the surface magnetic features. 
The process of absorption and damping of $p$-modes by an increasing 
number of rising flux tubes during the period of high activity explored 
by \citet{bogdan96} is qualitatively compatible with our results. On the other hand, if the TVP variations are only due to
geometrical effects, this parameter would probably show a better
correlation with the total area covered by the magnetic
structures than with just the number of them.
Although it is beyond the scope of this work, this should 
probably be further investigated  to get a better picture of the possible
sources of this phenomena.
A more quantitative work including separate observations
of the damping rate and amplitude variations  of individual 
modes \citep{Chaplin2000}
is certainly needed to be more conclusive. These inferences from
individual mode fits are potentially more informative but 
remains however less robust than those obtained
from the fit of the cross-correlation of the power spectra.

\section{Frequency dependence of the frequency shifts}\label{sec:nu-dep}

In the previous sections, the variations of the frequency-integrated
velocity power and  frequency shifts have been studied.
However, the time variation of the frequency shift is expected to
be different at different frequencies. 
Previous works  \citep[e.g.][]{libbrecht90,gubau92,chaplin98a} have shown 
that the modes at high frequency become more sensitive to the
solar cycle. The quality of our time-series and a new method to
fit all the spectra at once motivates us to determine the
frequency-dependence  of the frequency shift for low degrees.

The first method used is related again to the
fact that pairs of low-degree acoustic modes are equally 
spaced in the spectrum. Each spectra is divided in regions of 
$135$~$\mu$Hz containing a set of modes $\ell$=0,1,2 and 3. 
 Then, every region is cross-correlated with the corresponding
 region of the 
reference spectrum (corresponding to the solar activity minimum of 1986),
and  the method explained above (Sect.~\ref{sec:int_dnu})
 is used to calculate the 
frequency shift. Finally, the frequency shift of each region is 
fitted as a linear function of the integrated radio flux at 10.7 cm.  

In the second method proposed here, we try to fit together all the
spectra at once. We have established that the central frequencies
of the solar acoustic modes vary during the solar cycle and there is 
a strong linear correlation with any of the solar indices. 
So, in order to improve the statistics we can fit all the
spectra together introducing the frequency shift as a new parameter. 
Here, as well as in the first method, the radio flux was chosen 
because, according to Table~\ref{corr_dnu}, it leads to the best 
linear correlation coefficients for the central frequency variations but, 
as quoted before, this choice is not crucial as all indices present 
similar correlations.
We emphasize that, while the first method is faster, the second provides us
not only individual frequency shifts but also valuable mode parameters, 
i.e. mode resonant frequencies corrected for the solar-cycle effects. 

As the structure of the power spectrum is complicated
by the presence of the $11.57\mu$Hz sidebands, modes 
close in frequency must be fitted simultaneously in order to 
maintain the stability in the fits. 
Therefore adjacent $(n,\ell)$ and $(n-1,\ell+2)$ peaks are fitted together.  
The multiplet structure induced by the rotation and the temporal 
 sidebands for each  mode of the pair 
are also included in the model. 
If we label by $p$ the $60$~$\mu$Hz wide part of the spectrum including 
the adjacent $(n,\ell)$ and $(n-1,\ell+2)$ peaks and $\nu_p$ the mean frequency of the pair,
the model for this part  of the spectrum of each time-series $i$ can be 
expressed as:
\begin{eqnarray}
\label{m02}
M_p^i(\nu,\vec{a})= \sum_{k=-1}^{1} \beta_{|k|} \left[\sum_{m=-\ell}^{\ell} \frac{\alpha_{\mid m \mid}^{\ell} A_{n\ell} 
(\Gamma_{p}/2)^2}{(\nu-\nu_{n\ell m}^{ik})^2+ (\Gamma_p/2)^2} +
\nonumber \right.\\
\left. \sum_{m=-\ell-2}^{\ell+2} \frac{\alpha_{\mid m \mid}^{\ell+2} A_{(n-1)(\ell+2)} (\Gamma_{p}/2)^2}{
(\nu-\nu_{(n-1)(\ell+2)m}^{ik})^2 +(\Gamma_p/2)^2} \right]+
B_p,
\end{eqnarray}
with:

\begin{equation}
\label{nu_m02}
\nu_{n\ell m}^{ik} = \nu_{n\ell} + \delta\nu(\nu_{p})_{\ell,\ell+2}
 \cdot (F^i_{10} - F_{10}^o) + m \cdot s_{n\ell} + k_D \cdot k,
\end{equation} 
where:
\begin{itemize}
\item $\nu_{n \ell}$, is the central resonance frequency of the mode $(n,l)$ at solar minimum;
\item $\delta\nu(\nu_p)_{\ell,\ell+2}$, is the searched 
frequency shift per solar index unit, assumed to be the same for all
the components of the pair;
\item $\nu_p=(\nu_{n\ell}+\nu_{n-1,\ell+2})/2$ is the center of the $60\mu$Hz
wide interval of the spectrum fitted;
\item $s_{n \ell}$, is the synodic rotational splitting for a multiplet taken
to be constant for all of them and equal to $400$~nHz;
\item $F^{i}_{10}$, is the averaged radio flux
 for the given series ($F_{10}^o$ being the value 
at the solar cycle minimum);  
\item $B_p$, is a constant background level at the fitted frequency interval;
\item $A_{n \ell}$, is the power at the resonance for sectoral components $m$=$\ell$
\item $\alpha_{\mid m \mid}^\ell$, is a given ratio 
$A_{n \ell m}$/$A_{n \ell}$
 constrained to 
take the theoretical value, for an instrument without spatial resolution,
calculated by \citet{jcd89} for the special case of
resonant scattering observation using potassium line;
\item $\Gamma_p$, is the linewidth of the components of the multiplet
 assumed to be the same for all components of the two multiplets; and
\item $\beta_{|k|}$, is the ratio of the power of the sidebands ($|k|=1$)
to the
central peak ($|k|=0$).  These are directly related to the duty cycle
and can be estimated empirically. The duty cycles of the yearly time-series 
remaining close to their average value of $25\%$, this allows us to 
fix the amplitude of the sidebands
to $\beta_{|1|}$=0.5 for  $\beta_{0}$=1.
\end{itemize}

We assume that each $m$-component is well represented by a symmetric 
Lorentzian profile (even though \citet{sabri00} have found a slight 
asymmetry on them) and that the individual $m$-components 
are independent \citep[see][]{Foglizzo98}.
$B_p$, $A_{n \ell}$, $\Gamma_p$, $\nu_{n \ell}$ and 
$\delta\nu_{n \ell}$ are the parameters to be fitted 
(vector $\vec{a}$, hereafter). 
Notice that, in Sect.~\ref{sec:tvp}, it has been demonstrated that the 
TVP depends on the solar activity. This indicates that 
time variations may also occur for  amplitudes and linewidths.
Such variations have indeed been found recently 
from  BiSON \citep{Chaplin2000} for low degree
and GONG \citep{Komm2000} for higher degrees,
both  showing an increase of the linewidths and  
a decrease of the amplitudes leading to decrease of the TVP 
in agreement with our result. 
We therefore tried to parameterize amplitudes and linewidths
 as a function of
time in the way used for frequencies, but the fits
turned out to be very difficult and in many cases did not
converge. 

At higher frequencies ($\nu>3500$~$\mu$Hz) the peaks get wider, 
the width being greater than their rotational splittings 
($\Gamma_p$ $\gg$ $s_{n \ell}$). 
Moreover, the linewidths get so large that they become comparable to,
or bigger than, the small frequency separations between the pair $(\ell,n)$,
$(\ell+2,n-1)$.  Therefore the fit is  made at frequency intervals (labeled by $q$ hereafter)
of $165$~$\mu$Hz centered at $\nu_q$ and
  containing one pair of even modes $(n,0)(n-1,2)$ labeled hereafter by $p=1$ 
and one pair of odd modes $(n,1)(n-1,3)$ labeled hereafter by $p=2$. 
Only one Lorentzian is fitted  for even and another one for odd modes. 
Thus the model becomes:

\begin{equation}
\label{mn1}
M_q^i(\nu,\vec{a})=  \sum_{p=1}^{2}  \frac{ A_{pq} (\Gamma_{pq}/2)^2}{(\nu-\nu_{pq}^i)^2+
 (\Gamma_{pq}/2)^2} +B_q,
\end{equation}
with:
\begin{equation}
\label{mn2}
\nu_{pq}^i = \nu_{pq} + \delta\nu(\nu_q) \cdot (F^i_{10}-F_{10}^o).
\end{equation}
Here the parameters $\vec{a}$ to be fitted are: $B_q$, $A_{pq}$, 
$\Gamma_{pq}$, $\nu_{pq}$ and $\delta\nu(\nu_q)$. 

As \citet{Woodard_Phd} points out, in the case of observations without spatial resolution, the power spectrum of the 
solar $p$-mode oscillations is distributed  around the mean Lorentzian 
profiles with a $\chi^{2}$ probability distribution with two degrees of 
freedom. 
Consequently, the power spectrum appears as an erratic function where
an abundance of frequency fine structure can be found. 
For this type of statistics the joint probability function associated to 
the observed power spectrum $\vec{X}^i=\{X^i(\nu_j)\}_{j=1,N}$  
corresponding to the time-series $i$ is given by:
\begin{equation}
\label{dpf}
f^i(\vec{X}^i) = \prod_{j=1}^N \frac{1}{M^i(\nu_{j},\vec{a})} exp\left[-\frac{X^i(\nu_{j})}{M^i(\nu_{j},\vec{a})}\right],
\end{equation}
where N is the number of frequency bins in the interval considered (i.e. the 60~$\mu$Hz wide interval for $\nu<3.5$~mHz or the 165~$\mu$Hz wide interval 
for higher frequencies), and $M^i$ is given either by Eq.~(\ref{m02}) or Eq.~(\ref{mn1}) depending also on the frequency domain.

\begin{figure}
\resizebox{\hsize}{!}{\includegraphics{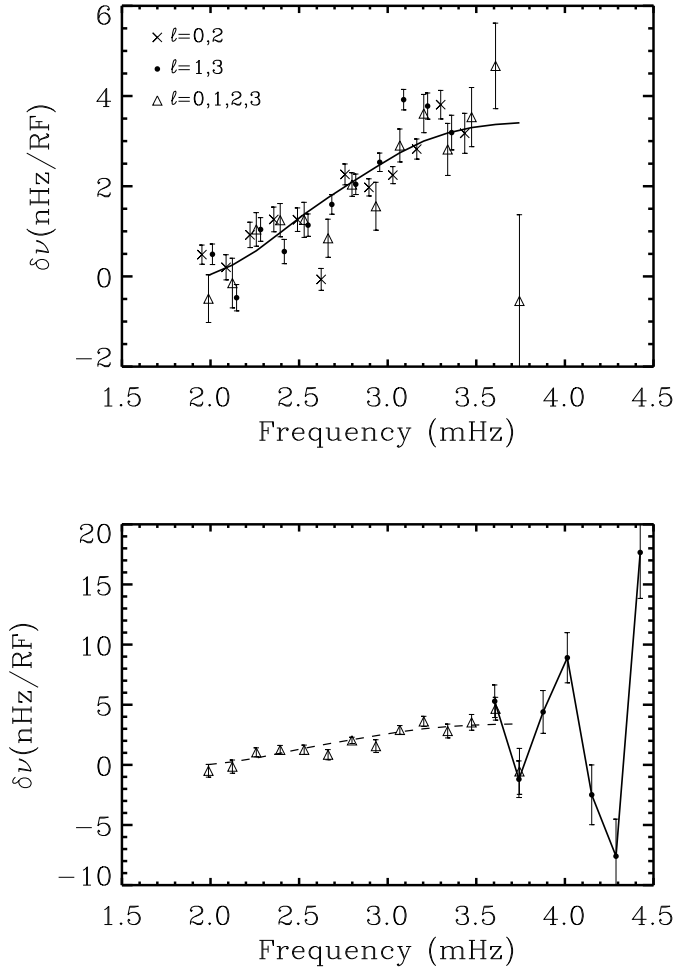}}
\caption{The slope of the frequency shift in its assumed linear 
dependence with the radio flux is plotted here as a function of 
the frequency for low-degree $p$-modes. The results for all 
modes (triangles) are obtained using  the first method described
in the text while the separate fits for odd and even 
degrees are  obtained using a simultaneous fit (Eqs.~(\ref{m02}), (\ref{nu_m02})).
The solid line is the best fit of the inverse mode mass calculated for 
the same set of modes. The bottom figure shows the results 
obtained at high frequency using Eq.~(\ref{mn1}).
The estimation of the frequency shifts corresponding to low degree
using the first method as well as the best fit of the inverse mode mass have 
been again plotted in the bottom figure as reference.
RF stands for radio flux units, namely $10^{-22}$ J/s/m${}^2$/Hz.}
\label{grad_shift_hn}
\end{figure}


The likelihood  for the 30 observed spectra is then given by 
the product of the individual likelihood function for each year. 
Thus, it can be written as:
\begin{equation}
\label{mle1}
L(\vec{a})= \prod_{i=1}^{30} {f^i(\vec{X}^i)}.
\end{equation}
We then look for the vector $\vec{a}$ that will maximize
the likelihood of the observed spectra according to our model.
For numerical reasons one minimizes $S$, defined as
the negative logarithm of the likelihood function,

\begin{equation}
\label{mle}
S(\vec{a})
          = \sum_{i=1}^{30} \sum_{j=1}^{N} \left[{\ln(M^i(\nu_j,\vec{a}))+
\frac{X^i(\nu_j)}{M^i(\nu_j,\vec{a})}}\right].
\end{equation}
To minimize this 
expression, we have used a modified Newton method \citep{numrec}. 
The initial guesses for the parameters are important to avoid local minima. 
In that aspect, the frequency shift appears to be the more sensitive 
parameter and the initial guess was taken from the results shown 
Table~\ref{slope_dnu} independently of the frequency range fitted.
In the case of some parameters (amplitude, noise and linewidth),
the natural logarithm of those have been fitted and not the 
parameters themselves.
Doing this, $S$  follows a normal distribution near the minimum and
the covariance matrix for the vector $\vec{a}$ can be approximated by
the inverse of the Hessian matrix found at the minimum of $S$. 
The uncertainties on each fitted parameter are therefore taken as the 
square roots of the diagonal elements of the inverted Hessian matrix.

The results are summarized in Fig.~\ref{grad_shift_hn}. 
In the top figure, the triangles represent the frequency shift 
integrated in bands of $135$~$\mu$Hz using the first method. 
The frequency dependence is quite clear, the frequency shift 
being close to zero at $2$~mHz and then increasing progressively  
with the frequency
to reach approximately $2$~nHz/RF  (where RF stands for radio flux units, namely $10^{-22}$J/s/m${}^2$/Hz)
at the center of the $p$-mode 
and a maximum of $4$~nHz/RF around $3.6$~mHz, where it drops fast. 
The solid line denotes the inverse mode mass extracted from the
solar model of \citet{morel97}.  
It has been averaged over the same regions in frequency and what 
we show represent the best fit to the data. 
The results corresponding to the simultaneous fits are also
shown in the same figure. 
The frequency shift for the even modes are represented by crosses while
black circles represent the odd ones. 
Both confirm the previous results showing a similar frequency dependence.
These results are also in  agreement with the analysis carried out
by \citet{chaplin98a} on BiSON data for frequency below $3.6$~mHz.
The mean $<\delta\nu(\nu)>$ in the 2.5-3.7mHz range can be estimated by
integrating the best fit of the inverse mode mass 
to our results divided by the length of the frequency range  i.e. 1.2mHz.
This leads to a value of $2.66$nHz/RF compatible with the slope $b$ reported
in Tab.~\ref{slope_dnu}  for the linear dependance between the integrated frequency shift and radio flux. 
A comparison of Eqs.~(\ref{eq:def1}) and (\ref{eq:fit}) gives the straightforward
correspondance between $b$ and the averaged frequency shift per radio flux
units.
We note however that the two estimates are not equivalent  due to
the variation of the $p$-mode energy across the five
minute band. The cross-correlation function weight
more the peak with higher amplitude located around 
$\sim2.9-3.0$mHz while the present integration  gives the same weight
to all the modes.

At high frequency, the first points corroborate the frequency
shift obtained using the first method, where a sharp downturn
is found. Then, at higher frequency, a large fluctuation appears.
A similar feature was found by \citet{gubau92} and also found 
for intermediate degree  by \citet{libbrecht90} in the analysis of BBSO data.
These authors observed that the sensitivity of the mode 
frequency shifts show similar sharp downturn located around $3.8$~mHz. 
\citet{chaplin98a} also reported (on their Fig.~4)
a sharp downturn with negative frequency shift of about $-10$~nHz/RF 
around  $4.3$~mHz but  they did not find the first downturn
found at $3.75$~mHz as in our analysis. The better sampling of the
Mark-I data set and the fact that this first downturn was obtained
with the two different methods we used for low and high frequencies 
make us confident in this result; moreover, this result is also found by 
\citet{gubau92}.

The frequency dependence of the frequency shift has been addressed
 from the theoretical point of view in different
works. The model developed by \citet{goldreich91} suggests that 
a combination of an  increase of 
the chromospheric temperature and a chromospheric resonance 
can be responsible for the 
sharp downturn at high frequency followed by an oscillation 
while the progressive increase 
in the five-minute band can be interpreted as an increase of the 
filling factor of the small scale photospheric magnetic fields.
On the other hand, \citet{jainroberts93,jainroberts96} argue that
the presence of a magnetic field in the chromosphere and a combination of 
temperature and magnetic field strength variations could, qualitatively,
explain the observed frequency dependence of the frequency shift at
both low and high frequencies. We notice that  both  models, those  of
\citet{goldreich91} and 
\citet{jainroberts96} present a \emph{wavelike} behavior at high frequency
qualitatively similar to the one found in our analysis. The particular
model of \citet{goldreich91} is even able to reproduce quantitatively 
this result, including the upturn around $4.5$~mHz but no evidence
has been found for the required 
chromospheric resonance \citep{woodard_librecht91} and \citet{kuhn98}
point out  that the change in the  photospheric magnetic field strength
needed in this model is much higher than the one obtained
from recent infrared splitting observations of the quiet region
field strengths or MDI magnetograms. 
Thus, \citet{kuhn98}  argues that photospheric magnetic 
fluctuations are unlikely to be responsible for the observed 
frequency shifts and proposes instead  
that turbulent pressure and mean solar atmosphere stratification
variations resulting from entropy perturbations through the solar 
cycle may be the dominant process affecting $p$-mode frequencies.
In this model, the associated temperature fluctuations may originate
from near the base of the convection zone but  in order 
to explore those possibilities and locate the different possible
perturbations one needs to invert the even splitting coefficients  
\citep[see][]{Dziembowski00}
and track any fluctuation in the sound speed inversion using long term
observations. We intend to extend 
our work in that direction  
using the  observations of both low and  
intermediate degrees from LOWL and Mark-I.

\begin{figure}[t]
\resizebox{\hsize}{!}{\includegraphics{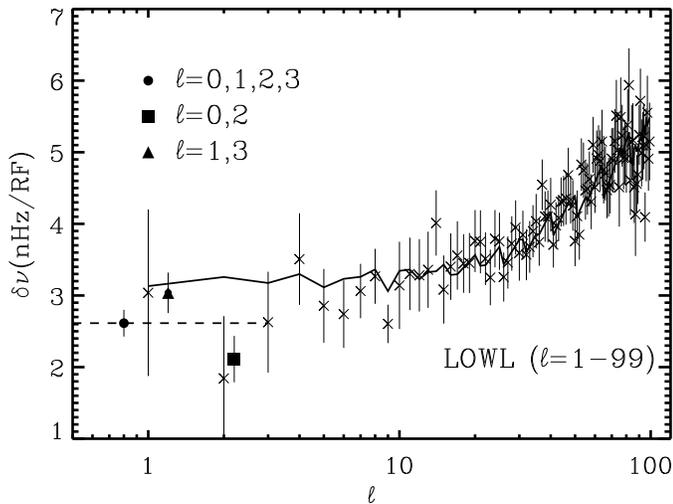}}
\caption{The cross shows the normalized frequency shift for 
$\ell$=1-99 integrated between $2.5$ and $3.7$~mHz using LOWL
data, and the solid line denotes the best fit of the inverse 
mode mass for those modes.
We have also plotted the results corresponding to very low degree using
Mark-I data (Tab~\ref{slope_dnu}): 
the dashed horizontal bar is for $\ell$=0,1,2,3; the triangle 
for $\ell$=1,3 is placed close to $\ell$=1 because this component
 gives the main contribution; and the square for $\ell$=0,2 
has been arbitrarily placed close to $\ell$=2.}
\label{dnu_flux2}
\end{figure}


\section{Analysis of the $\ell$-dependence of the frequency shifts}\label{sec:l-dep}

The LOWL instrument \citep{LOWL}, located in Mauna Loa, Hawa\"\i,
is a Doppler imager based on a Potassium Magneto-Optical Filter,
and it has been collecting solar observations for more
than six years. With the installation of a similar experiment 
at the Observatorio del Teide, a new network called ECHO 
(Experiment for Coordinated
Helioseismic Observations) intends to continue the solar observations
for a complete solar cycle \citep{ECHO} with an increased duty cycle.
Recently, six years of data have been re-analyzed through a 
new pipeline producing mode parameters for low and intermediate degrees 
\citep{jimenez-reyes01b}. 
We have used the mode frequencies given by this analysis to
compare the $\ell$- and frequency-dependence of the frequency shift at 
low and intermediate degrees.

Figure~\ref{dnu_flux2} shows the normalized frequency shift using
LOWL observations for $\ell$=1 up to 99. 
It has been performed using those modes between $2.5$ and $3.7$~mHz, 
as we did for low degree.
The inverse mode mass, which was calculated as well for each one of 
the modes fitted and then averaged in the same way, follows 
remarkably well the results. 
The figure shows as well the $\ell$-dependence of low degree $p$-mode 
frequency shifts for Mark-I found in Sect.~\ref{sec:int_dnu}.
The general trend confirms the $\ell$-dependence of the
frequency shift, the sensitivity at high degree ($\ell$=100) being
almost twice to that of low-degree. We notice that the LOWL $\ell$=2
and Mark-I $\ell=$0,2 are in good agreement and significantly
lower then the inverse mode mass curve. However the LOWL data error bars in
$\delta\nu$ for $\ell=1$ and $\ell=2$ overlap significantly and the 
small value for $\ell=0,2$ could either be due to particularly small
$\delta\nu$ in $\ell=2$ but also in $\ell=0$. It will therefore be important 
to check this results with independent observations in the future.
 This is important because if a 
geometrical effect purely related to the integrated disk
 measurements affect the even modes measured by Mark-I, this 
 should not be seen for the  LOWL resolved measurements. If confirmed,
 this lower
value of the frequency shift for $\ell$=2 may therefore have 
a  physical origin and be the signature of
a localized perturbation.
\begin{figure}[t]
\resizebox{\hsize}{!}{\includegraphics{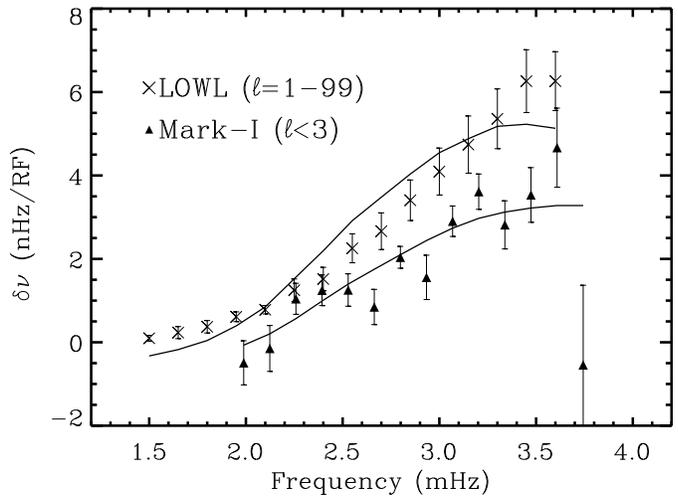}}
\caption{Normalized frequency dependence of LOWL data averaged for
modes from $\ell$=1 up to 99 in intervals of $150$~$\mu$Hz. 
The results for very low degree shown in Figure~\ref{grad_shift_hn} 
are also plotted. The solid lines represent the best fit of
the inverse mode mass to both data sets.}
\label{dnu_flux1}
\end{figure}

Figure~\ref{dnu_flux1} reproduces the frequency dependence for low degree
  shown in  Fig.~\ref{grad_shift_hn}  together with the frequency
shift averaged for $\ell$'s between 1 up to 99 in
intervals of $150$~$\mu$Hz obtained from LOWL data. 
Again, the best fit to the inverse mode mass is shown. 
The ratio between the two slopes 
in the inverse mode mass fits is $0.74$ and is nearly equal to 
the ratio between the mean mode mass calculated for both mode sets ($0.75$)
 showing that the $\ell$-dependence of the frequency shift is 
again well described by the $\ell$-dependence of the inverse mode mass.
However, although the sensitivity is higher 
for higher degree  modes (LOWL data), 
the fit is worse than for low degree (Mark-I data); moreover, 
in this later case,  while roughly consistent with the sizes of the error bars,the scatter  seems to be organized as an oscillation 
on top of the  fitted line. 
The higher sensitivity of the frequency shift at high degree seems to be 
in contradiction with what \citet{regulo94} pointed out, 
but in agreement with recent analysis of \citet{chaplin98a}.
While all results point towards the 
existence of a perturbation confined close to the surface, there is still 
no convincing evidence of another cause that could exist deeper down as 
suggested by \citet{regulo94}. 
On the other hand, the oscillation, also pointed out by \citet{gubau92},
that seems to be present in the results for low degree,
has the same period ($\approx$ 400-450 $\mu$Hz) that the 
one that is clearly found at higher frequencies 
(see Fig.~\ref{grad_shift_hn} bottom) and deserves 
further attention and confirmation.

Finally, we notice that the frequency dependence of the frequency shift 
obtained from the low degree modes of LOWL data is in very good agreement
with the result plotted here using Mark-I data. This, added to the fact
that the 6 years of LOWL data were covered also by Mark-I observations,
 make us very confident in the validity of our analysis of
the $\ell$-dependence using both instruments.

\section{Conclusion}
An analysis of the low-degree p-mode frequency shifts over more than a solar 
cycle has been carried out. 
Time dependence of the resonant frequencies and the total velocity power have 
been studied with several, old and new, methods, yielding a correlation 
and an anti-correlation with the solar activity cycle respectively 
which confirms previous results. 
Moreover, quantitative parameters have been introduced which yield the 
sensitivities of such variations with respect to a given activity index, 
in our case the radio flux at 10.7 cm. 
The existence of a hysteresis behavior between both parameters 
seems to imply different direct causes for them.

Also, the frequency shifts have been obtained as a function of frequency 
and degree. These studies make use of data taken from two experiments 
(Mark-I and LOWL) which confirm each other results, when they coincide 
(at very low $\ell$ modes), and complement their findings otherwise. 
Indeed, the main source for the variation of the frequencies with the cycle is 
located near the surface while a secondary, deeper rooted source, seems to be 
very weak if it exists at all. 
Interesting details such as the oscillation found for very low-$\ell$ p-modes 
and the surprisingly small frequency shift for $\ell$ even modes need 
further investigation.

The results obtained and methods used in this work will also be useful 
in the analysis of very long time-series covering more than half the solar 
activity cycle, since both resonant frequencies and, to a lesser extent, 
amplitudes vary cyclically with time, providing wrong results if only data 
taken at parts of the cycle are used. In the following paper 
(second part of this work) we will use these methods to minimize the 
effects of solar activity on these parameters in annual power spectra, 
allowing an average of all data thus improving the signal to noise ratio. 
Using this power spectra we will be able to give a new estimation of the 
solar rotational splitting, which combined with LOWL data, will be used to 
infer the solar rotation close to the core.

 Finally we note that
the hysteresis between frequencies of odd and even modes or between mode
frequencies and magnetic flux
is not addressed in this work based on yearly time series. These studies
are however of considerable importance
\citep[see e.g.][]{fmi00} but would require shorter
time series and  better
data (e.g. higher duty cycle) in order to reliably confirm previous
analysis. This should probably be addressed
in the future using long term observations of low degree modes from the
ground based
networks BiSON and IRIS++ ( i.e. IRIS + LOWL +Mark-I, \citet{salabert01}).

\begin{acknowledgements}
We are deeply thankful to the all members, past and present, of the
helio\-seismo\-logy group at the IAC for doing Mark-I ob\-ser\-va\-tions
and maintenance.
The use of Birmingham University resonant scattering spectrophotometer
at Observatorio del Teide is also deeply acknowledged.
We are extremely grateful to Tom Bogdan and Mausumi Dikpati for 
useful discussions
and additional comments. T. Corbard acknowledges support from NASA grant
S-92678-F and PPARC grant PPA/A/S/2000/00171. The High Altitude Observatory
of the National Center for Atmospheric Research is sponsored by the
National Science Foundation.
\end{acknowledgements}


\end{document}